\documentclass[pre,aps,twocolumn,showpacs]{revtex4}
\usepackage{graphicx,epsf,subfigure,amsbsy,amsmath}

\usepackage{notoccite}

\usepackage{amssymb}
\usepackage{amsfonts}
\usepackage{mathrsfs}

\begin{document}

\title{Topological origins of a bi-parameter periodicity hub for the R\"{o}ssler attractor}

\author{Timothy D. Jones$^1$}

\affiliation{$^1$Physics Department, Drexel University, Philadelphia,
  Pennsylvania 19104, USA}
\date{\today}

\begin{abstract}
We explore the dynamical and topological characteristics of
the R\"ossler system that 
lead to the existence of 
a periodicity hub and nested spiral in codimension-2 parameter space.  
We find that the hub shape is a 
consequence of conjugacy between the R\"ossler system and
unimodal maps in the two-branch region of the parameter space.  The
nested spiral structure is a consequence of a topological feature
of the R\"ossler system that has not been noted previously.  We outline
this mechanism and detail the spiral transition for the symbolic
sequence of orbits up to period seven.
\end{abstract}

\pacs{XXXXXXX}
\maketitle

Recent work has focused attention on the appearance of nested spirals and
so called ``shrimp'' within the codimension-2 parameter space of a number of 
dissipative systems.  First noted by Bonatto and Gallas in 2008 \cite{Bonatto2008}, these hubs have been the 
focus of number of investigations since \cite{Barrio2009,Gallas2010,Ramirez-Avila2010,Freire2010, Barrio2011z, Shilnikov2011, Vitolo2011}.

\begin{figure}[h]
\includegraphics[angle=0,width=7.0cm]{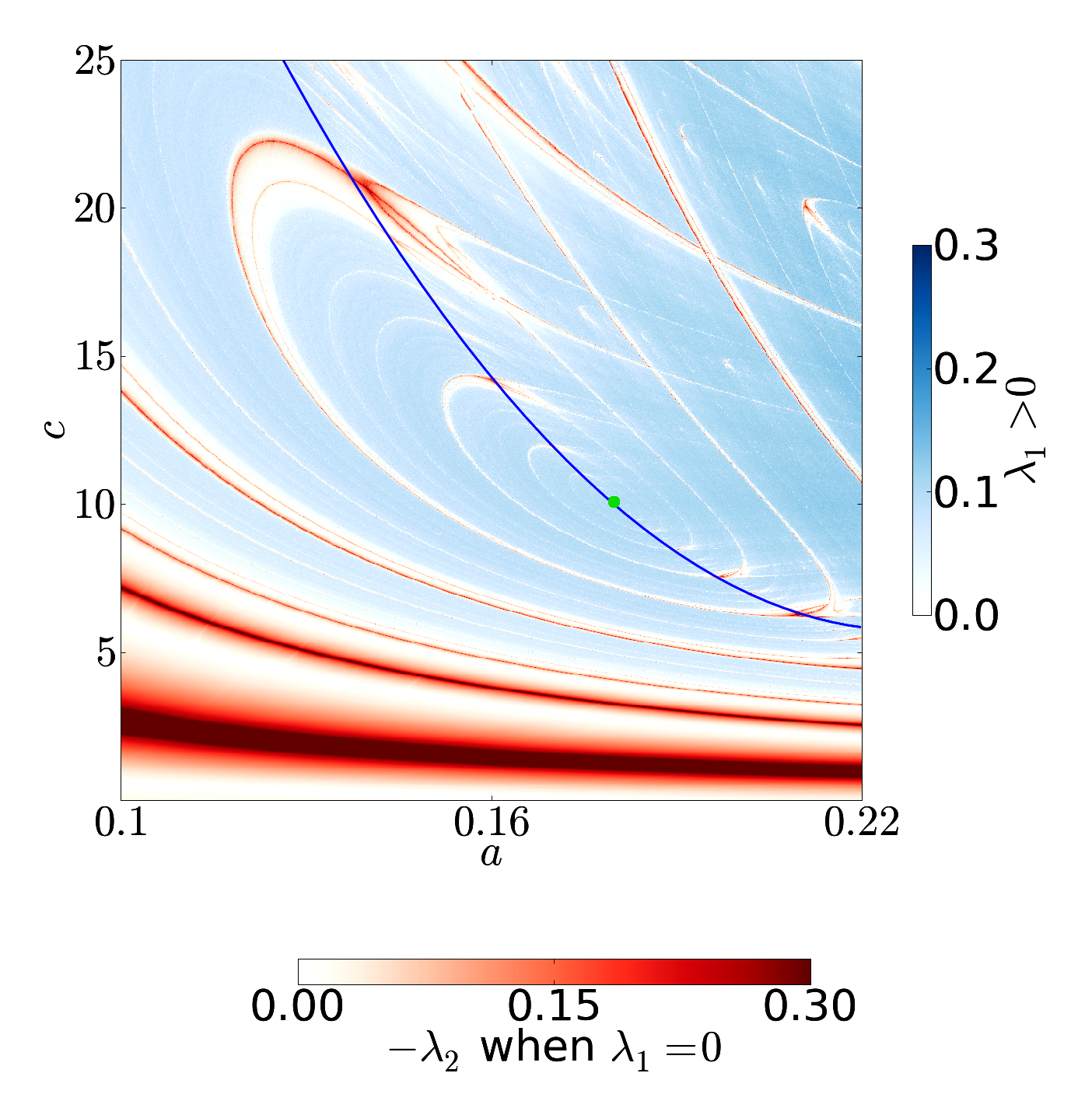}
\caption{Lyapunov diagram for the R\"ossler system, close up of primary spiral hub.  Green center dot (color online)
indicates the center of the spiral hub.  Dark blue line indicates a best-fit curve for the partition
of the primary hub by the appearance of ``shrimp''.  Blue coloration reflects the intensity of the first Lyapunov exponent and corresponds
to chaotic regions,
red coloring reflects that of the negative of the second exponent when $\lambda_1 =0$ and corresponds to stable regions. 
Lyapunov exponents were calculated on a 1000x1000 grid using the LESNLS software package \cite{Dieci2011}.}
\label{fig2:hubby}
\end{figure}

The occurrence of these hubs is easily seen in a mapping of the global Lyapunov exponents in 
codimension-2 parameter space (Figure \ref{fig2:hubby}).  We focus on a hub found in the R\"ossler system.
The hubs for this system have been studied for a number of different parameter 
values \cite{Barrio2009}.  We chose to work with the parameter space of the 
R\"ossler system where $b=0.2$ and $(a,c)$ vary.  It can be shown that the
shape of this hub is robust with a variation of $b$, drifting rightwards
and slightly downwards in the $a,b$ parameter plane as $b$ increases from $0.2$
through $2.0$. 

The equations for the R\"ossler system \cite{Rossler1976} are,   
\begin{equation}
\label{eq20}
\dot{x}= -y-z, \ \  \dot{y} = x+ay, \ \ \dot{z}=b+z\left( {x-c} \right)
\end{equation}
While this system was first designed as a model for chaos, with no practical use in mind, 
numerous dynamical systems in fields such as chemistry, electronics, biology, and lasers have
been found to exhibit behavior conjugate to that of the R\"ossler system \cite{gilmore2011topology}.

In Figure \ref{fig2:hubby}, we draw an approximate best-fit curve linking the centers of so called ``shrimp'' structures along a line
which corresponds to a transition from ``spiral'' (the return-map of the R\"ossler attractor has two branches)
 to ``screw-like'' structure in the R\"ossler attractor, which corresponds
to the branched-manifold describing the R\"ossler attractor obtaining a third branch and the return-map appearing bimodal with
two critical points.  We call this line the Topological Transition Line and write this particular line as $TTL^2_3$ to indicate
the transition from needing a branched-manifold with two branches to one with three branches as we move in parameter space from 
left to right \cite{Gilmore1998a}.
In agreement with previous estimates \cite{Barrio2011z}, we find the center of the 
primary spiral hub to be numerically located at approximately $(a,c)=(0.1798,10.3084)$ which
corresponds to a homoclinic bifurcation point, the only such point along $TTL^2_3$ \cite{Bonatto2008}. 

We first wish to understand the smooth structure left of the $TTL^2_3$,
and we do so by assuming topological conjugacy between the R\"ossler system and unimodal
maps for judiciously selected slices of parameter space.  That is, while the return map 
for the R\"ossler attractor appears conjugate with unimodal maps left of $TTL^2_3$, we 
seek a more specific conjugacy--that of the spectrum of stable periodic orbits.  For
this purpose, we will compare the R\"ossler system with the most widely studied unimodal
map.  The Logistic Map is described by the equation 
$L(x_n,r)=x_{n+1}\equiv rx_n\left( 1-x_n\right)$.
A method of analytically detecting 
periodic windows in the general class of Logistic type maps was introduced by the seminal work
of Metropolis et al. in 1973 \cite{Metropolis1973}, and rediscovered by Jensen and Myers in 1985 \cite{Jensen1985,Eidson1986}.  
In order to locate
windows of stability, we 
can study the iterations of the critical point. 
Whenever one of these iterations of the critical point intersects the critical point, we are
assured of the existence of a super-stable periodic orbit which corresponds to one of the windows
of stability seen in the Logistic Map's bifurcation diagram.  
These iterations of the critical
point ``scar'' the bifurcation diagram as continuous ``caustics'' and stand out due to a higher concentration of ``hits''
around these points.
A further advantage of the caustic method is that this method also describes the symbolic encoding of the
super-stable orbits \cite{Metropolis1973}.  Similar caustics are seen in bifurcation diagrams taken
for the R\"ossler system, but not necessarily with the same patterns as in the Logistic Map. 

The apparent conjugacy between the R\"ossler bifurcation
diagram, and that of the Logistic Map, has not been commonly noted (see Letellier 1995 \cite{Letellier1995a} for an exception).
This is likely due to the fact that this similarity can only be found if one takes a slice through parameter
space that is properly oriented with regards to periodicity conjugacy requirements between the maps.  
In order to judiciously select a slice in parameter space
through which we can assume topological conjugacy with the spectrum of stable periodic orbits in unimodal dynamics,
 we note that the main
conjugacy requirements of the Logistic Map will be
\begin{small}
\begin{itemize}
\item{Unimodal structure.}
\item{An initial period-1 orbit that undergoes a period doubling cascade and then transitions to chaos upon reaching the 
Feigenbaum point.}
\item{The attractor evolves until it makes contact with a fixed point.}
\end{itemize}
\end{small}
The R\"ossler system matches these requirements in a specific region,
\begin{small}
\begin{itemize}
\item{Unimodal structure:  due to the highly dissipative nature of the R\"ossler system, its return map has a unimodal 
nature in the regions left of $TTL^2_3$. 
}
\item{The R\"ossler system undergoes an Andronov-Hopf Bifurcation, which leads to a period-1 limit cycle that undergoes a period-doubling
cascade as we get closer to $TTL^2_3$ in parameter space.}
\item{The R\"ossler system has a homoclinic bifurcation curve (a thin parabola shape), the tip of which lies on $TTL^2_3$ \cite{Bonatto2008}.  
The attractor makes contact with its fixed point in the form of a homoclinic orbit at this point.}
\end{itemize}
\end{small}

We treat the R\"ossler system as if it were fully topologically conjugate to the Logistic Map for any line in the parameter
space which begins at the Andronov-Hopf Bifurcation and terminates at the homoclinic point on the $TTL^2_3$  
where the R\"ossler attractor first makes contact
with its central fixed point.  This results in the shape of the hub left of $TTL^2_3$ tracing out the shape of the Andronov-Hopf bifurcation curve. 

\begin{figure}[h]
\includegraphics[angle=0,width=6.0cm]{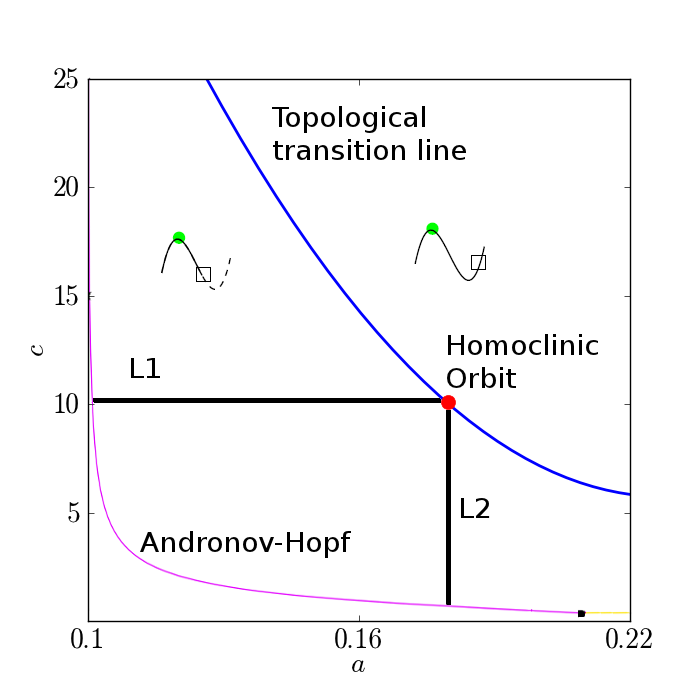}
\caption{Parameter space diagram with Andronov-Hopf curve calculated using PyCONT/AUTO \cite{ClewleyRH2007,auto07p}.  
Lines $L1$ and $L2$ originate from the
period-1 limit cycle created after the Andronov-Hopf bifurcation of the R\"ossler attractor and terminate at 
a Homoclinic bifurcation point on the curve marking the topological
transition from two to three branches in the branched manifold.  The order of 
stable periodic orbits
seen along these lines are conjugate to those seen in the Logistic map up to at least period seven and likely for 
much higher periods.   We exclude the continuation of the homoclinic bifurcation curve
into the three-branch region for simplicity.}
\label{hubby:fig3d}
\end{figure}

In Figure \ref{hubby:fig3d}, we draw an example
of two such slices of parameter space that will fulfill these conjugacy requirements, $L1$ and $L2$.  We find that their
bifurcation diagram is qualitatively similar to that of the Logistic map, and furthermore, we tested this assumption by numerically 
recreating the first seven caustics in these bifurcation diagrams.  At least up to period seven and likely so for higher periods, 
the order of periodic orbits in the 
R\"ossler system as we move along these lines is conjugate with that of the Logistic Map.  
\begin{figure}[h]
\includegraphics[angle=0,width=8.4cm]{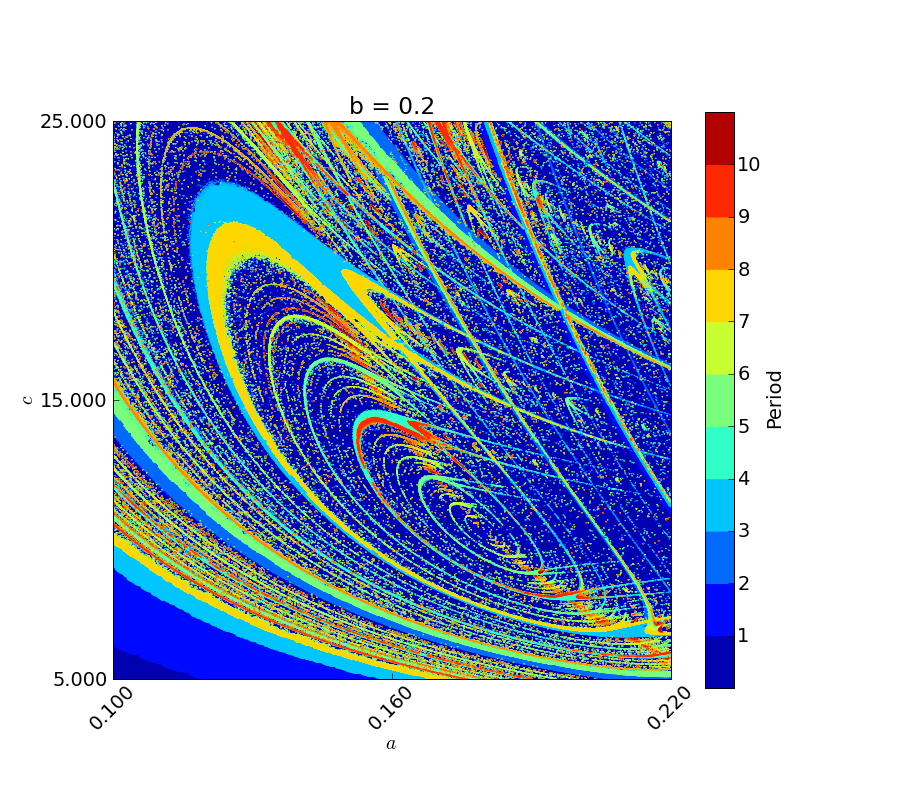}
\caption{Periodicity calculations for the codimensional-2 parameter space of the R\"ossler attractor, for $b=0.2$, $a\in(0.1,0.22)$, 
$c\in(5.0,25.0)$. 
Darkest blue corresponds to divergent or chaotic behavior.}
\label{periods_wide:hubby}
\end{figure}
To further test these assumptions, we partitioned parameter space around the spiral hub into a 5000 by 5000 grid and applied a 
periodicity detection algorithm for each
point.  We color code the periods, up to period 11, producing an extensive periodicity graph in Figure \ref{periods_wide:hubby}
which shows that the periodicity is structured as predicted.  

Finally, we can see that by continuity, this ordering of periods also determines the order of periods of the shrimp structures 
along the $TTL^2_3$.  
At the center of these shrimps, one finds
a degenerate super-stable state which corresponds to the fact that at these points two intersecting parabola-shaped regions of super-stable periodicity
contain both critical points simultaneously \cite{Gallas1994}.  Careful consideration of the conditions at the intersection of the 
topological regions will help us understand why the shrimp line up along the transition curve. These stable regions 
represent iterations of at least one critical point of the return-map.  
The first two iterations of the critical point mark the borders of the range of possible iterations.
As we pass through $TTL^2_3$ in phase space from left to right, 
the R\"ossler's state-space return map grows a third branch.  At $TTL^2_3$,
the edge of the second branch is precisely the second critical point
of the bimodal map in the branch-3 region. 
\begin{figure}[h]
\includegraphics[angle=0,width=8.5cm]{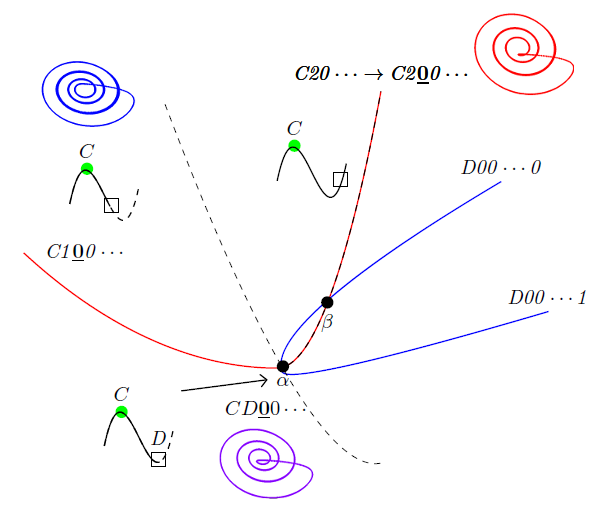}
\caption{Anatomy of a generic ``mutant shrimp'' along the largest spiral.
The symbolic string describing these orbits will be dominated by zeros.  The center of the shrimp structure
corresponds to a doubly-superstable orbit which includes both critical points.  Each parabola corresponds
to a variation of this orbit in which only one of the critical points is on the orbit.  
Due to the topological mechanism responsible for the periodicity transition, this shrimp has one tail which originates
as a lower period and transitions to higher period before crossing $TTL^2_3$.}
\label{mutant_shrimp}
\end{figure}

The shrimp
along the $TTL^2_3$ and above the focal point behave as we would
expect in the bimodal regime \cite{Gallas1994}.  One of the ``tails'' of each of the shrimp points downward back
towards the hub.  This ``tail'' joins a shrimp below the focal point
which has a different period.  We call these ``mutant shrimp'' and have detected
the periodicity transition to happen very close to $TTL^2_3$ (Figures \ref{periods_wide:hubby} and \ref{mutant_shrimp}).

Directly related to these structures is the question of the source of the nested periodicity spirals.
These spirals are due to a previously unnoticed topological feature of the R\"ossler system. 
\begin{figure}[h]
\includegraphics[angle=0,width=5.5cm]{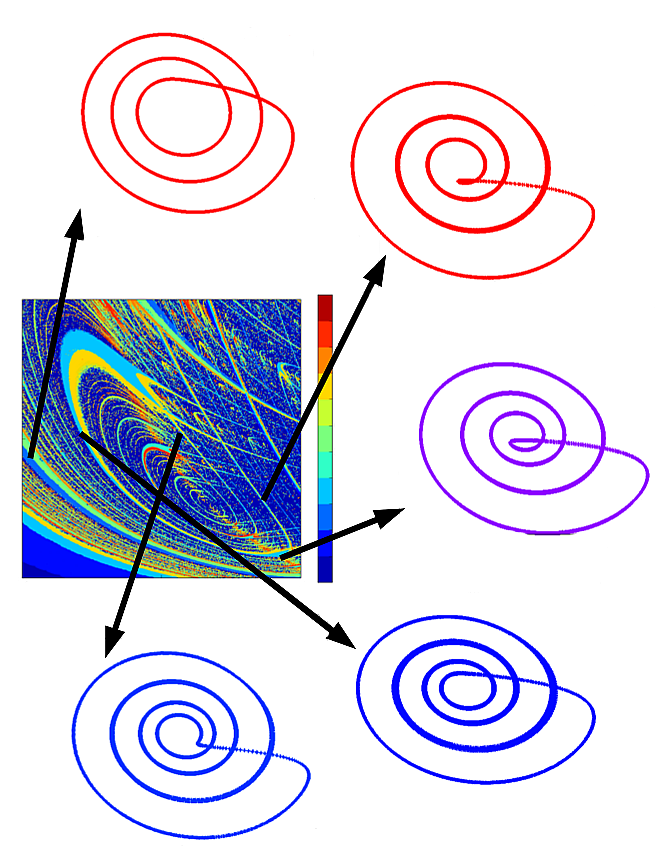}
\caption{
As we trace the periodicity spiral around in parameter space, clockwise from lower to higher period, the 
third branch of branched-manifold in state space is reinjected on the plane of the attractor at an angle
that varies clockwise around the fixed-point, merging with the second branch as it crosses the $TTL^2_3$ from
right to left. }
\label{periods_shrimp4:hubby3}
\end{figure}
The branched-manifold describing the R\"ossler system has been typically assumed to have a standard shape in
which all branches are redeposited onto the main plane of the attractor at approximately the same position.  We find that this
is not the case for a small range of values in codimension-2 parameter space.  In particular, as we move around
one of the spiral features in the hub in parameter space, we find that in state space, the position at which
the third branch is deposited on the main plane of the attractor rotates around the fixed point.  Figure \ref{periods_shrimp4:hubby3}
shows this clearly for the spiral transition from period three to period four.   
As we follow a stable periodic
orbit clockwise around the central spiral hub from lower periodicity to higher periodicity, the angle of deposit
also rotates clockwise around the central dynamical fixed point in state space.  Like a long chain being deposited in a circular way 
on a flat surface, the number of turns of the chain on the flat surface increases as we follow the 
periodicity regions as they spiral in towards the center of the spiral hub.  As we cross $TTL^2_3$ following
these spirals, the third branch ``collapses'' back onto the second branch, and the cycle starts over.  This results in the
transition between lower to higher period happening very rapidly near the shrimp along the bottom section of $TTL^2_3$
below the homoclinic point. 

Following these spirals inwards, we find that the new points on an orbit
are deposited on the inner-most region of the 
attractor, 
winding the attractor closer and closer to its fixed point until it makes
full contact at the homoclinic bifurcation.
\begin{figure}[h]
\includegraphics[angle=0,width=8.2cm]{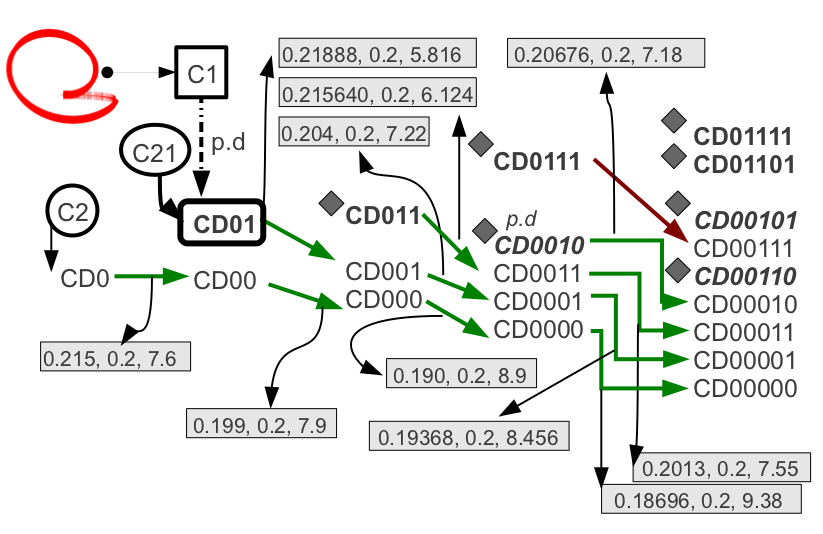}
\caption{
Symbol sequences in the nested spiral hub up to period seven.  
The orbits listed in each column are in the order that they appear in unimodal
mappings \cite{Gilmore1998a}.  Green arrows indicate orderings which have been matched both
by looking at the structure in state space, and by matching the symbol sequence of the corresponding
transitions in parameter space.  The transition $CD0111 \rightarrow CD00111$ has
only been confirmed by visually inspecting the attractor.
Orbits in bold originate
from isolated shrimp off of $TTL^2_3$ within the branch-3 side.  The attached gray boxes indicate
the approximate position of the corresponding mutant shrimp in parameter space.  
}
\label{sumbol_diagram}
\end{figure}
Knowing this new mechanism, we can predict the symbolic sequence of orbits along a spiral. 
We must have that the critical point associated with the unimodal 
map on the branch-2 side of $TTL^2_3$, which we label $C$, is always part of the symbolic sequence along this spiral.
This is due to the fact that the spiral region is continuous and crosses over the branch-2 side where
the bimodal critical point, which we label $D$, is excluded. 
Thus both connected periods $p$ and 
$p+1$ will begin with symbol $C$.  
For example, we know that the period-four orbit $C100$ must transition into an
orbit $C1X00$ after passing through the sequence $C200$ and returning to the $TTL^2_3$ since the third
branch lays the new point on the orbit,
where $X$ is either $0$ or $1$.  
Since we know that the new branch is deposited 
on the inner region of the main plane of the attractor, this new symbol will be a $0$, and so we have
the transition from $C100 \rightarrow C200 \rightarrow C1000$.  We have traced the trains for a number of orbits (Figure
\ref{sumbol_diagram}) which follow this general algorithm, though we note that higher periods will have
a more complicated algorithm.  We also find that 
the orbit $CD0$
is connected with an orbit $C2$ inside the branch-3 region, and $CD01$ has been found to be
connected to an orbit $C21$ also from this region.  Finally, we note that $CD01$ is the period-double
of the orbit $C1$ which itself is connected to a period-one orbit in the branch-3 region which transitions 
to the orbit $C1$ by the same mechanism outlined in this report.  The shape of this period-1 orbit is
outlined in the upper left of Figure \ref{sumbol_diagram} as it nears transition.

Orbits that can not originate within a spiral, because the would-be preceding orbit does not exist, instead
originate from one of the isolated shrimp in the branch-3 side of parameter space, though they themselves
are then the beginning point of a new spiral, resulting in a nested set of spirals.  

\section*{Acknowledgments}  
This work is supported in part by the U.S. National Science Foundation under grant PHY-0754081.  The author is grateful
to Drs. Vallieres and Yuan for generous use of their computational clusters, and to Dr. Robert Gilmore for encouragement 
and useful suggestions. 


\end{document}